\documentstyle[amssymb,aps,preprint,12pt]{revtex}
\tightenlines

\begin{document}
\title{NMR comparative study of PbMg$_{1/3}$Nb$_{2/3}$O$_{3}$ and PbSc$_{1/2}$Nb$%
_{1/2}$O$_{3}$ local structure.}
\author{V. V. Laguta, M. D. Glinchuk, S. N. Nokhrin, and I. P. Bykov,}
\address{Institute for Problem of Material Sciences, Ukrainian Academy of\\
Sciences,\\
Krjijanovskogo 3, 03142 Kiev, Ukraine}
\author{R. Blinc, A. Gregorovi\v {c}, and B. Zalar}
\address{Jo\v {z}ef Stefan Institute, P. O. Box 3000, 1001 Ljubljana, Slovenia}
\date{\today }
\maketitle
\draft

\begin{abstract}
The $^{93}$Nb and $^{45}$Sc NMR spectra in PbSc$_{1/2}$Nb$_{1/2}$O$_{3}$
(PSN) and PbMg$_{1/3}$Nb$_{2/3}$O$_{3}$ (PMN) disordered relaxor
ferroelectrics at the temperature T $>$ T$_{0}$ (T$_{0}$ is the temperature
of the dielectric susceptibility maximum) have been studied. Spectra
analysis was performed on the base both of the analytical description of NMR
lines shapes, allowing for homogeneous and inhomogeneous broadening related
to a random distribution of the electric field gradients and numerical Monte
Carlo method taking into account electric field gradients originated from
random distribution of Mg, Sc and Nb ions (which may be shifted or not) over
B-type cation sites.

The observed 1/2 $\longleftrightarrow $ -1/2 transition spectrum both of the 
$^{93}$Nb and $^{45}$Sc nuclei in the PSN was shown to contain a narrow (3-4
kHz) almost isotropic part and a broad strongly anisotropic part. These two
components of NMR spectra are related to 1:1 Sc/Nb ordered and
compositionally disordered regions of the crystal, respectively. It was
shown that in the disordered regions Sc$^{3+}$, Nb$^{5+}$and O$^{2-}$ ions
are shifted from their cubic lattice sites at one of three possible
directions: $<100>$, $<110>$ or $<111>$.

In PMN the NMR spectrum of $^{93}$Nb contains practically only the broad
component. The portion of unbroadened spectrum that may correspond to ideal
1:2 regions accounts only for 1-2 percent of the total integral intensity.
No evidance was obtained about existence of the 1:1 regions in PMN. The NMR
data demonstrate that in PMN the cubic symmetry at T $>$ T$_{0}$ is locally
broken due to ions shifts similar to that in disordered PSN. The values of
the ion shifts were estimated in the point charges point dipoles
approximation of the electric field gradients calculation both in the PSN
and PMN.
\end{abstract}

\pacs{77.84.Dy, 76.30.Fc}

$\cdot $\narrowtext

\section{Introduction}

Relaxor ferroelectrics \cite{Smol54,Cross87} have been attracting
considerable attention in recent years due to their unusual physical
behavior. Lead magnesium niobate, PbMg$_{1/3}$Nb$_{2/3}$O$_{3}$ (PMN), is
probably the best investigated perovskite relaxor crystal. Many of novel
models which try to describe the physical properties of PMN are grounded on
the assumption of the existence of a 1:1 Mg/Nb ordered microregions at least
in 1/3 of the volume of a crystal. On the other hand, the latest x-ray and
neutron diffraction measurements performed on the PbSc$_{1/2}$Nb$_{1/2}$O$%
_{3}$ (PSN) with different degree of cation ordering show that ferroelectric
phase transition undergoes in both ordered (1:1 Sc/Nb) and completely
disordered samples \cite{Malibert97}. It might mean that long-range ordering
can not be suppressed completely by compositional (chemical) disorder.
Moreover, a large shifts of all type of ions from cubic regular positions in
PMN as well as in PSN already at T $>$ T$_{0}$ (T$_{0}$ represents the
temperature of dielectric constant maximum) were obtained from the computer
treatment of the x-ray and neutron diffraction peaks. Due to this, the phase
transition even in PSN becomes incomprehensible because, as it follows from
x-ray data, at T $>$ T$_{0}$, for instance, Pb$^{2+}$ ions are shifted away
of about 0.027 nm with equal probability along one of the directions $<100>$%
, $<110>$ or $<111>$. This consideration left unclear, however, if these ion
shifts are the static or dynamical ones.

Another difficulty in the interpretation of the x-ray and neutron
diffraction data in relaxors is related to amorphous character of observed
diffraction peaks even at T $<<$ T$_{0}$. These methods provide an averaged
view of the structure resulting in the cubic (PMN) or close to cubic
symmetry (PSN). Broadening of diffraction peaks is caused by the
compositional disorder of these materials. Thus, local methods of magnetic
resonance and, in particular, Nuclear Magnetic Resonance (NMR) can be
extremely useful in the study of such disordered systems because in NMR
experiment the nuclei are sensitive to their environment only at a distance
less then 1 - 2 nm. It provides also possibility to distinguish between the
static and dynamic ion shifts that is practically unattainable in x-ray
diffraction methods.

In our previous papers (see, for instance, Ref. \cite{Glin}) we have already
studied $^{93}$Nb and $^{45}$Sc NMR spectra both in the PMN and PSN. In
particular, it was shown that at T $>>$ T$_{0}$ NMR spectra of $^{93}$Nb are
mainly determined by the electric-field-gradient contribution originated
from disorder in Nb/Mg(Sc) lattice sites. From NMR spectra analysis
evidences were obtained about the existence of ordered and disordered
microregions in both materials. Unfortunately, earlier measurements were
carried out mainly at a low Larmor frequency ($\nu _{L}\simeq $ 49 MHz) so a
bigger part of the $^{93}$Nb spectrum belonging even to 1/2 $%
\longleftrightarrow $ -1/2 transition remained invisible and could not be
analysed. More detailed $^{93}$Nb and $^{45}$Sc NMR spectra were recently
presented in Ref.\cite{Blinc} and \cite{psn}. However, only the temperature
dependence of the second moment of the narrow subset of NMR lines has been
analysed. A broad background observed in the spectrum was attributed to
unresolved satellite transitions and it was not analysed.

The purpose of the present paper is to study local structure both of PMN and
PSN at temperatures T $>$ T$_0$ on the base of $^{93}$Nb and $^{45}$Sc NMR
measurements performed at Larmor frequency $\nu _L\simeq $ 93 MHz. We
believe that complete 1/2 $\longleftrightarrow $ -1/2 transition spectrum
can be observed in such conditions. Because one of studied materials (PSN)
belongs to a 1:1 system, where the 1:1 Sc/Nb order really exists it should
be thus interesting to see what are the common features and what are the
differences in NMR spectra of these two types of relaxors.

Description of the NMR line shape was performed on the base of analytical
formulas, allowing for both homogenious and inhomogenious mechanisms of line
broadening. As a second method of NMR spectra analysis we used numerical
Monte Carlo simulation, which allowed to take into account random
distribution of Mg, Nb and Sc ions (shifted away or not) over B-type sites
of AB$^{\shortmid }$B$^{\shortparallel }$O$_{3}$ perovskite lattice.

The structural data obtained are compared with analogous data resulting from
x-ray diffraction methods. Finally, the problem of 1:1 Mg/Nb ordered
microregions in PMN is discussed as well.

\section{Samples and experimental details}

The measurements were carried out on single-crystal samples of PMN and PSN
grown on a seed crystal from solution in melt. Samples had a good optical
quality. The sample dimensions of PSN were (4x4x4) mm$^3$, and PMN - up to
(8x8x8) mm$^3$ with the surfaces parallel to crystallographic (001) planes.

The NMR measurements were performed in a 9.2 T superconducting magnet at $%
^{45}$Sc Larmor frequency $\nu _L$ = 92.3 MHz (I = 7/2) and $^{93}$Nb $\nu
_L $ = 92.9 MHz (I = 9/2). Temperature was stabilized in a continuous-flow
cryostat with an accuracy of about 0.1 K. The inhomogeneously broadened
spectra of $^{93}$Nb and $^{45}$Sc ( 1/2 $\longleftrightarrow $ -1/2
transition) have been obtained using the Fourie transform of the spin echo
after a 90$_x$ - $\tau $ - 90$_y$ - $\tau $ - acquisition pulse sequence.
The width of the 90$^0$ pulse was 2.1 $\mu s$ and 2.4 $\mu s$ for $^{93}$Nb
and $^{45}$Sc resonances, respectively. A time delay $\tau $ was usually 40 $%
\mu s$ for both of nuclei.

\section{Experimental results and their interpretation}

\subsection{$^{93}$Nb and $^{45}$Sc NMR spectra}

Measurements were performed at T = 400 - 450 K because we were interested
mainly in the local structure of PMN and PSN in high temperature phase, i.
e., at T $>$ T$_{0}$ (T$_{0}\eqsim $ 355 K and 310 K for PSN and PMN,
respectively). As an example in Fig. \ref{fig1} we presented $^{45}$Sc NMR
spectra measured for three characteristic crystal orientations in the
external magnetic field: B II [001], [011] and [111] . The characteristic
feature of the spectrum for B II [001] and [111] is its strong asymmetry
while for B II [011] the spectrum is almost symmetric. Similar spectra of $%
^{93}$Nb in the PSN and PMN are shown in Figs. \ref{fig2} and \ref{fig3},
respectively. Let us discuss first the Sc and Nb spectra in the PSN.

Both the Sc and Nb spectra have approximately similar structure and both
contain a narrow component of the width of (3 - 4) kHz and a wide component
that is different for the Sc and Nb nuclei. The narrow part of spectra is
well described by a Gaussian and is completely isotropic. On the contrary,
the wide part is essentially anisotropic and can not be precisely described
by a simple function. In rough approximation, however, it can be also
described by a Gaussian (see in Ref. \cite{psn}). In such approach the ratio
of integral intensity of narrow and wide parts of the spectrum is
approximately equal (40:60) and does not depend on the crystal orientation.
Note that due to large width of the $^{93}$Nb spectrum its wide component
has very small intensity and thus its anisotropy manifests itself much less
than that in the $^{45}$Sc spectrum. However measurement of whole the $^{93}$%
Nb spectrum in PSN, using the sweep of the irradiation frequency, indeed
showed that the Nb spectrum shape is very close to that for $^{45}$Sc
(insert to Fig. \ref{fig2}).

$^{93}$Nb spectrum in PMN is somewhat complicated. In this spectrum we can
conditionally separate three components: (1) low-intensity isotropic line
with the width of 3 kHz; (2) spectrum around $\nu _L$ with the width of 15 -
18 kHz; (3) wide and strong anisotropic component like that in PSN but more
intensive. The narrow part with $\Delta \nu $ = 3 kHz accounts only for
about 1.5 - 2 percent of the total integral intensity of the spectrum. As
can be seen from Fig. \ref{fig3}, beside the narrowest component the whole
spectrum is strongly anisotropic.

The anisotropy of spectra as well as asymmetric form with respect to $\nu
_{L}$ for both of $^{93}$Nb and $^{45}$Sc nuclei and in both the PMN and PSN
points out that given spectra belong to 1/2 $\longleftrightarrow $ -1/2
transition the frequency of which is shifted by the second order quadrupolar
contributions \cite{Abraham}. This quadrupole contribution exists only for
nuclei with spin I $>$ 1/2 and is related to the interaction of the nuclear
quadrupole moment with the electric field gradient (EFG). Large values of
quadrupole moments of $^{93}$Nb and $^{45}$Sc niclei ($eQ=-0.28e\cdot
10^{-28}$ m$^{2}$ and $-0.22e\cdot 10^{-28}$ m$^{2}$ for $^{93}$Nb and $%
^{45} $Sc, respectively) usually result in predominance of the quadrupole
contribution to NMR frequency shift with respect to other possible sources,
as chemical shift and magnetic dipole-dipole interaction. Therefore we can
draw conclusion that main features of $^{93}$Nb and $^{45}$Sc NMR spectra in
both the PMN and PSN are related to characteristic features of their EFG,
that depend on a lattice ions charges and their positions, i.e. local
structure of crystals.

\subsection{Analysis of $^{45}$Sc and $^{93}$Nb spectra in PSN}

The integral intensity ratio (40:60) obtained above for the narrow and wide
parts of the $^{45}$Sc and $^{93}$Nb NMR spectra in PSN is in agreement with
the Sc/Nb degree of ordering which varied in the studed crystals in the
range of 30 - 40 percent. So, it can be assumed that narrow and wide parts
of the spectra belong to the Sc and Nb nuclei situated in the ordered 1:1
and disordered regions of the crystal, respectively. Actually, in the
ordered regions Sc and Nb ions are surrounded by a symmetrical ion
configuration. Therefore in such sites the electric-field-gradients and
anisotropic chemical shift must be close to zero and we can expect a NMR
line broadened only by the magnetic dipole-dipole interaction. Calculation
of this dipole-dipole width gives the value of about 3 kHz supporting the
origin (ordered 1:1 regions) of narrow lines in $^{45}$Sc and $^{93}$Nb NMR
spectra.

Evidently, the wide part of NMR spectra belongs to resonances from nuclei
located in the disordered regions of the crystal, where random distribution
of nonequal charges B$^{\prime }$/B$^{\prime \prime }$ ions produce
electric-field-gradients. It follows from recent x-ray and neutron
diffraction data \cite{Malibert97} that in these regions the lead and oxygen
ions may be shifted away from their regular cubic positions of about 0.01 to
0.035 nm, which gives an additional contribution to the EFG tensor
components. Of course, due to disorder in the ions positions we can expect
broad distribution in values of the EFG tensor components which manifestes
in inhomogeneous broadening of NMR lines.

The resonance frequency shift related to quadrupole mechanism for the
considered case of $\pm $ 1/2 transition is proportional to the squared EFG.
It can be written as follows

\begin{equation}
\nu _{1/2}^{(2)}=\frac{\nu _{Q}^{2}}{6\nu _{L}}\left( I\left( I+1\right) -%
\frac{3}{4}\right) f_{\eta }(\theta ,\varphi ),  \label{eq1}
\end{equation}

where $\nu _{Q}=\frac{3e^{2}qQ}{2I\left( 2I-1\right) h},eq=V_{zz},\eta =%
\frac{V_{xx}-V_{yy}}{V_{zz}}.$

The function $\ f_{\eta }(\theta ,\varphi )$ in Eq.(\ref{eq1}) describes the
dependence of frequency shift on mutual orientations of external magnetic
field and EFG tensor principal axes, $\nu _{L}$ is the Larmor frequency,
x,y,z form the pricipal EFG tensor reference frame.

Angular variation of the wide part of NMR spectra can be qualitatively
understood if to assume presence of an axial EFG pointed at $<$001$>$ cubic
directions. In this case angular part of expression (\ref{eq1}) takes a
simple form:

\begin{equation}
f(\theta ,\varphi )=-\frac{3}{8}\left( 1-\cos ^{2}\theta \right) \left(
9\cos ^{2}\theta -1\right) .  \label{eq2}
\end{equation}

Expected resonances for all six $<$001$>$ directions of the EFG tensor are
depicted as hystograms in Fig. \ref{fig1}. In this our simple interpretation
we have to take into account that due to large disorder in the ions
positions EFG values become randomly distributed that leads to strong
inhomogeneous broadening of NMR lines. As a result, only the shift of the
center of gravity is visible in the experimental spectrum under crystal
rotation. Obviously, that because of nonlinear relation between resonance
frequency shift and quadrupole frequancy inhomogeneously broadened NMR line
acquires essentially asymmetrical form, which besides depends on orientation
of a crystal in external magnetic field. For such complex spectrum its
decomposition on simple Gaussian components and the analysis of their
angular dependence can not give true values of quadrupole frequencies.

In what follows, we introduce the distribution of largest EFG component
which leads to the distribution of quadrupole frequencies $\nu _{Q}$. Let us
suppose that this distribution function can be expressed in Gaussian form,
i. e.

\begin{equation}
F(\nu _{Q})=\frac{1}{\sqrt{2\pi }\Delta }\exp (-\frac{(\nu _{Q}-\nu
_{Q}^{0})^{2}}{2\Delta ^{2}}),  \label{eq3}
\end{equation}

where $\nu _Q^0$ is the mean value of the quadrupole frequency and $\Delta $
is the width of its distribution.

Allowing for the nonlinear relation between $\nu $ and $\nu _Q$ (see Eq.(\ref
{eq1})), namely

\begin{equation}
\nu =\alpha \nu _{Q}^{2},\text{ }\alpha \equiv \frac{1}{6\nu _{L}}\left(
I\left( I+1\right) -\frac{3}{4}\right) f(\theta ,\varphi ),  \label{eq4}
\end{equation}

the shape of inhomogeneously broadened line can be written as (Ref. \cite
{GlKon})

\begin{equation}
P(\nu )=\sum\limits_{j=1}^2\frac{F(\nu _Q=\nu _{Qj})}{\mid \frac{d\nu }{d\nu
_Q}\mid _{\nu _Q=\nu _{Qj}}},\text{ }\nu _{Q_{1,2}}=\pm \sqrt{\frac \nu %
\alpha .}  \label{eq5}
\end{equation}

Substitution of Eq.(\ref{eq3}) into Eq.(\ref{eq5}) gives

\begin{equation}
P(\nu )=\frac 1{^{\sqrt{2\pi \alpha \nu }\Delta }}\exp (-\frac{\nu +\alpha
(\nu _Q^0)^2}{2\alpha \Delta ^2})\cosh \frac \alpha {\Delta ^2}\sqrt{\frac %
\nu \alpha .}  \label{eq6}
\end{equation}

To take into account the additional homogeneous broadening mechanism with
Gaussian lineshape (dipole-dipole interactions) one has to make the
integration

\begin{equation}
I(\nu )=\frac{1}{\sqrt{2\pi }\delta }\int_{-\infty }^{\infty }P(\nu
^{^{\prime }})\exp \left[ \frac{-\ln 2(\nu -\nu ^{\prime })^{2}}{\delta ^{2}}%
\right] d\nu ^{\prime },  \label{eq7}
\end{equation}

where $\delta $ being a half of the Gaussian width.

One can see that Eqs. (\ref{eq6}-\ref{eq7}) give an asymmetrical lineshape
which depends also on the parameter $\alpha (\theta ,\varphi )$ that takes
into account the mutual orientations of EFG and magnetic field B$_0$.

One can expect that in disordered materials there are microregions with
different $^{\prime \prime }$degree of order$^{\prime \prime }$ and
different directions (and mean values) of the ion displacements. Obviously,
in such a case the NMR line should be the composition of the lines $^{\prime
\prime }$stemming$^{\prime \prime }$ from these microregions, i.e. 
\begin{equation}
I_{q}\left( \nu \right) =\sum\limits_{i}I_{qi}\left( \nu \right)  \label{eq8}
\end{equation}

where subscript $^{\prime \prime }$i$^{\prime \prime }$ numerates the
microregions.

The results of simulation on the base of Eqs.(\ref{eq6}-\ref{eq8}) of the $%
^{45}Sc$ NMR line shape for ${\bf B}\Vert \lbrack 001]$ are presented by
solid line in Fig. \ref{fig4}(a) and (b). Figure \ref{fig4}(a) shows the
line shape for the case of EFG with largest principal axis points in $%
\left\langle 001\right\rangle $ directions only. The contribution from
ordered regions, where $\nu _{Q}=0$ was included also. The best fit has been
achieved for $\mid \nu _{Q}^{0}\mid $ = 755 kHz and $\Delta $ = 210 kHz. The
contribution of homogeneous broadening was estimated as $\delta $ = 2 kHz.
One can see that calculated NMR line has a strongly asymmetrical shape,
which fits very good the right $^{\prime \prime }$shoulder$^{\prime \prime }$
of the measured spectrum. The fitting of left $^{\prime \prime }$shoulder$%
^{\prime \prime }$ is not so good. This can be expected because for $%
\left\langle 001\right\rangle $ orientation of EFG only a positive
quadrupole shift of resonance frequency exists (see Eq.(4) when $\theta
=90^{0};0^{0}$). Therefore, to fit the left shoulder of the observed
spectrum the other orientations of EFG axes are required. The negative shift
is possible for EFG axes pointing along $\left\langle 011\right\rangle $ and 
$\left\langle 111\right\rangle $ directions. The reason for this type local
distortions can be related to substitutional disorder in the relaxors.

Really, the probability of configurations with $k$ ions of B$^{^{\prime }}$
type in the nearest neighbourhoods of B$^{^{\prime }}$ or B$^{^{\prime
\prime }}$ ions can be written as

\begin{equation}
P_6^k=\frac{6!y^k(1-y)^{6-k}}{k!(6-k)!}\text{, }y=1/3  \label{eq9}
\end{equation}

One can see that $P_6^k$ for $k=1,2,3$ are close to each other, while the
probability of having $k$ = 6 or 0 is much smaller than the most probable
value $P_m$ = 0.328 corresponding to $k$ = 2. It is easy to see that
configurations with $k=1,2,3$ can be the response of the three
afore-mentioned lattice distortions.

Finally, the experimental line shape of $^{45}$Sc was well fitted by the
following theoretical curve 
\begin{equation}
I_{q}\left( \nu \right) =C_{0}I_{ord}\left( \nu \right)
+C_{1}I_{<001>}\left( \nu \right) +C_{2}I_{\left\langle 110\right\rangle
}\left( \nu \right) +C_{3}I_{<111>}\left( \nu \right)  \label{eq10}
\end{equation}

with C$_0\simeq 0.3,C_1\simeq 0.45,C_2\simeq 0.11,C_3\simeq 0.14$; the
subscripts denote the direction of the EFG axes. The result of $^{45}$Sc
complete spectrum simulation is presented in Fig. \ref{fig4}(b).

The obtained set of fitting parameters ($\nu _{Q}$ and $\Delta $ are
presented in Table \ref{table1}) permits to describe theoretically the NMR
line shapes measured for any other orientation of magnetic field. In
particular in Fig. \ref{fig4}(c) and (d) the observed and calculated line
shapes for ${\bf B}\Vert \lbrack 011]$ and ${\bf B}\Vert \lbrack 111]$ are
shown. One can see the good agreement between measured and calculated line
shapes.

Similar simulations performed for $^{93}$Nb NMR spectra are presented in
Fig. \ref{fig5} and corresponding fitting parameters are listed in Table \ref
{table1}. The agreement between measured and calculated spectra seems also
good, however, due to small intensity of the wide part of the spectrum
obtained parameters may contain significantly bigger miscalculation than
that for the $^{45}$Sc spectrum.

\subsection{Analysis of $^{93}$Nb NMR spectra in PMN}

It can be seen from Fig. \ref{fig3} that practically all the $^{93}$Nb
spectrum in PMN is strongly broadened in comparison with the dipole-dipole
width that is equal to 2.8 - 3.6 kHz (as in the case of PSN). The portion of
the spectrum unbroadened by quadrupole mechanism or by anisotropic chemical
shift accounts only for 1 - 2 percent of the total intensity. It is clear
that the Nb nuclei, which give the contribution into this narrow spectrum,
locate at the positions close to the regular cubic ones. Possible origin of
these crystal regions will be discussed in the next section.

The wide, main part of the $^{93}$Nb spectrum is related to the disordered,
locally noncubic, regions of the crystal , as in the case of PSN. Contrary
to PSN, the portion of such regions in PMN accounts for 98$\%$ of the total
volume of the crystal.

$^{93}$Nb spectrum for disordered PMN was fitted assuming local distortions
along $<$001$>$, $<$110$>$ or $<$111$>$ directions, i.e. as in the case of
the PSN by Eq. (\ref{eq8}). Because the procedure of the simulation is
completely similar to that described for PSN we confined here only
presentation in Table \ref{table1} of spectral parameters derived.
Compareing spectral data for disordered PSN and PMN one can emphesize a
similarity in the symmetry and sometimes even in the values of electric
field gradients at Nb positions. One can see also that dispertion in PMN is
larger than in PSN. This is in agreement with the fact that in PMN local
lattice distortions should be much larger due to bigger difference in the
ionic charges of B$^{\prime }$ and B$^{\prime \prime }$ cations.

Summarizing, three main conclusions should be made.

(i) In the ordered regions, which account up to 30$\%$ in the PSN and 2-3$\%$
in the PMN of the total crystal volume, lattice ions occupy cubic lattice
positions where electric field gradients are close to zero.

(ii) In the disordered regions there are non-zero EFG along pure cubic
directions ($<$001$>$, $<$110$>$ or $<$111$>$). Because both the PSN and PMN
has in average cubic symmetry at T $>$ T$_{0}$, these non-zero EFG can be
related to randomly distributed unequal charges of B-type cations and/or
random shifts of the ions from their positions in the cubic lattice. In this
case the essential part of ions ($\sim $ 30-55$\%$) experiences shift along $%
<$001$>$ axes.

(iii) EFG tensor components are significantly distributed around their mean
values that manifests in broad asymetric NMR line shape.

\subsection{Numerical calculation of the EFG tensor and NMR line shape}

In the previous section on the basis of analytical calculation of the shape
of inhomogeneously broadened NMR line the mean values of quadrupole coupling
constants in Sc and Nb cation positions as well as their dispertions were
obtained. The method applied by us has allowed to take into account only
distribution of the quadrupole frequencies ($\nu _{Q}$), but the presence of
random deviations of EFG axes from their most probable directions, i.e the
possible fluctuations of $f_{\eta }(\theta ,\varphi )$ (see Eq. (\ref{eq1}))
remained discounted. Note that this effect is mainly related to a distant
ion contributions. Ignoring of these fluctuations has resulted in too much
high values of homogeneous width, which was accordingly 4 kHz and 8 kHz for
the $^{45}$Sc and $^{93}$Nb in PSN and 18 kHz for the $^{93}$Nb in PMN. Let$%
^{\prime }$s remind, that dipole width for both the $^{45}$Sc and $^{93}$Nb
accounts only 2.8-3.2 kHz.

The second reason on which we have undertaken numerical calculation of EFG
is related to attempt to define ions displacements, which could be
responsible for symmetry and magnitude of an EFG in both relaxor systems.
Taking into account complexity and substantially uncertainty of local
structure of these compaunds we used the simple model of point charge point
dipole, which, however, took into account effects related to partial
covalency of bonds in the BO$_{6}$ octahedron as well as electronic
polarizibility of ions. Obviously due to limitations of the point charge
model the obtained ions shifts should be considered only as an estimated
ones, the degree of which reliability depends on correctness (luck) of a
choice of such empirical parameters of ions as their effective charges and
polarizibility.

In the point charge point dipole approximation the electric field gradient
components at the site of the nucleus are given by

\begin{equation}
V_{ij}=(1-\gamma _\infty )\sum_k\frac{\partial ^2}{\partial x_i\partial x_j}%
\left[ \frac{q_k}{r_k}+\frac{r_{ik}\mu _{ik}}{r_k^3}\right] ,  \label{eq11}
\end{equation}

where $q_{k}$ and $\mu _{k}$ are the electric point chages and dipole moment
on the k-th ion. $r_{k}$ is the distance between the k-th ion and the
observation point, and $\gamma _{\infty }$ is the anti-shielding parameter 
\cite{Das56}

For the conviencity three contributions into EFG were detached in the whole
sum (\ref{eq11}). The first contribution ($V_{ij}^{(1)}$) was related to
random distribution of charges of B$^{\prime }$/B$^{\prime \prime }$ ions at
lattice points of ideal ABO$_{3}$ perovskite structure. Contributions of
ions at a distance up to five lattice constants (2 nm) were summed up.
However, usually the distance up to 1.2 - 1.5 nm was sufficient due to the
rapid decay of V$_{ij}$ with the distance (see Eq. \ref{eq11}). In the
calculations the B type ions were randomly distributed at lattice sites with
the help of random number generator or they were located in the 1:1 or 1:2
ordered regions. The main condition for any ion configuration was
electroneutrality, i. e. the stoichiometric (in average 1:1 for PSN and 1:2
for PMN) composition in the calculated region was conserved.

As an example, the result of the calculation of largest component $V_{zz}$
as a function of polar angle $\theta $ for the PMN is presented in Fig. \ref
{fig6}. Here we took effective valence charges of Mg$^{2+}$ and Nb$^{5+}$ as
one haft of their formal charges. It is seen that $V_{zz}$ lies close or
along cubic directions, but it is strongly distributed in both the magnitude
and orientation. The mean value $\mid \overline{V_{zz}}\mid =0.78\ast
10^{16} $ V/cm$^{2}$ ($\frac{e^{2}qQ}{h}\simeq 15MHz$) is much less than
that expected from coupling constant measured. Thus, this contribution to
the EFG can not be responsible for wide component of the $^{93}$Nb NMR line.
It can give only some contribution to homogenious broadening of NMR line
because it leads to distribution of axes directions of the total EFG. Note
that in PbSc$_{1/2}$Nb$_{1/2}$O$_{3}$ this part of EFG is neglectly small in
comparison even with the dipole-dipole width, that is in good agreement with
smaller difference between charges of Sc$^{3+}$ and Nb$^{5+}$ and, in
addition, 1:1 type relaxors have more symmetrical distribution of B$^{\prime
}$/B$^{\prime \prime }$ ions.

The second contribution (V$_{ij}^{(2)}$) into EFG took account for the
shifts of B$^{\prime }$/B$^{\prime \prime }$ ions themselves relatively to
the nearest six oxygen ions. Possible shifts of Pb$^{2+}$ ions were not
taken into account due to their smaller (10-15 times) contribution into EFG
for the same values of the shifts.

In a general case for arbitrary B type ion shift with respect to the center
of the undistorted oxygen octahedron, the following expressions for EFG
tensor components were obtained :

\begin{eqnarray}
V_{ij}^{(2)} &=&-\frac{42\cdot z_O^{val}}{b^5}d_id_j,\text{ }i\neq j 
\nonumber \\
V_{xx}^{(2)} &=&\frac{21\cdot z_O^{val}}{b^5}(2d_x^2-d_z^2-d_y^2)
\label{eq12} \\
V_{yy}^{(2)} &=&\frac{21\cdot z_O^{val}}{b^5}(2d_y^2-d_z^2-d_x^2)  \nonumber
\\
V_{zz}^{(2)} &=&\frac{21\cdot z_O^{val}}{b^5}(2d_z^2-d_x^2-d_y^2),  \nonumber
\end{eqnarray}

where (d$_x$, d$_y$, d$_z$) are vector components of the Nb or Sc shift, $%
z_O^{val}$ is the effective valence charge of the oxygen, $b=a/2$ ($a$ is
the lattice constant). For the simplisity the $z_O^{val}$ was taken as an
avarage value over all six oxygen ions. Its value may lay in the range
between $-1.63\mid e\mid $ (BaTiO$_3$ \cite{A}) and $-0.85\mid e\mid $ (KNbO$%
_3$ \cite{B}).

The last term in Eq. (\ref{eq11}) is related to the induced electric dipole
moments $\mu _{i}$. Assuming that only electronic polarization of the
nearest six oxygens gives marked contribution to the EFG (the other ions in
the unit cell and the next nearest oxygens give contributions that are an
order of magnitude smaller), the dipole contibution is:

\begin{eqnarray}
V_{xx} &=&V_{yy}=-\frac{42\mu }{b^5}d  \label{eq13} \\
V_{zz} &=&\frac{84\mu }{b^5}d,  \nonumber
\end{eqnarray}

where $\mu $ is assumed to be in the direction of the $Z$ axis of the field
gradient.

In the case of $<$110$>$ and $<$111$>$ symmetry of lattice distortions
nondiagonal components of the EFG tensor have to be also appeared. They are
given by expression

\begin{equation}
V_{ij}=-\frac{84\mu _i}{b^5}d_j,\text{ }i\neq j.  \label{eq14}
\end{equation}

Electronic dipole moment dependes on ionic polarizability and value and
direction of ions shifts and it can be usually calculated using knowledge of
the polarization and internal electric field. Obviously for disordered
relaxors such a way is not suitable because here is only local polarization
and large uncertainty in the ions positions. Therefore, for the estimation
of $V_{ij}$ we applied a method of effective dipole moments. In particular,
it is known, that in perovskite ferroelectrics the spontaneous polarization
can be well described by the Born effective charges:

\begin{equation}
P=\frac{\mid e\mid }\Omega \sum_sZ_s^{*}d_s,  \label{eq15}
\end{equation}

where $\Omega $ is the cell volume, and the Born effective charges $Z^{\ast
} $ are 0.82(K), 9.13(Nb), -6.58(O$_{1}$), -1.68(O$_{2}$) in tetragonal
phase of KNbO$_{3}$ \cite{C} and 3.87(Pb), 7.07(Ti), -5.71(O$_{1}$), -2.51(O$%
_{2}$) in PbTiO$_{3}$ \cite{D}. The high values $Z^{\ast }$ indicate on
larger contribution of electronic dipole part to whole polarization in the
comparison with rigid ionic part, that should be as a rule in high
polarizable perovskite ferroelectrics \cite{Zhong94}.

Taking into account a special role of Pb ions, namely their strong
hybridization with oxygen ions, we used in our calculations the effective
charges derived in PbTiO$_{3}$. For simplisity the same avaraged value of $%
Z_{O}^{\ast }=-4\mid e\mid $ was taken for evaluation of the electronic
dipole moment $\mu _{i}=Z_{O}^{\ast }\ast d_{i}$ in expressions (\ref{eq13}-%
\ref{eq14}). One can easy see, that under these conditions expressions (\ref
{eq12} ) and (\ref{eq13}-\ref{eq14}) become similar, thus in the calculation
of the whole dipole contribution to the EFG one can use only its electronic
part, because $Z_{O}^{\ast }>>Z_{O}^{val}$. Note that similar results were
obtained in KNbO$_{3}$\cite{Hewitt61} and BaTiO$_{3}$\cite{Viskov66}.

Taking into account that the ion shifts can have random values due to the
disorder of the Mg, Nb and Sc positions, the ion shift distribution function
was assumed in the Gaussian form:

\begin{equation}
f(d)=\frac 1{\sigma (2\pi )^{1/2}}\exp \left( -\frac{(d-d_0)^2}{2\sigma ^2}%
\right) ,  \label{eq16}
\end{equation}

where $d_0$ is mean value of the shift and $\sigma $ is its dispersion.

Complete NMR line shape was calculated taking into account
electric-field-gradient contribution, magnetic dipole-dipole interaction, as
well as chemical shift mechanisms. Since the contribution of the last
mechanism was much less with respect to the quadrupole one it manifests
itself only in the central narrow part of the spectrum. Its contribution was
accounted by the renormalization of dipole-dipole linewidth. NMR spectrum
intensity $I(\nu )$ was obtained by summarizing the contributions of the NMR
lines from a large number of clusters in which the B type ions could be
shifted in one of three directions: $<$001$>$, $<$011$>$, $<$111$>$ with
their resonance frequencies given by Eq. (\ref{eq1}).

\begin{equation}
I(\nu )=\sum\limits_{n=1}^{N}\exp \left[ \frac{(-\ln 2)\cdot (\nu _{n}-\nu
)^{2}}{\delta ^{2}}\right] .  \label{eq17}
\end{equation}

Here, $\nu _{n}$ is the resonance frequency for nucleus in the center of
n-th cluster, $\delta $ is a half of the dipole-dipole width and $N$ is the
total number of clusters (usually, $N$ = 15000 - 20000).

In the calculation of second-order quadrupole shifts for $^{93}$Nb the
parameter $\gamma _\infty $ $=-16$ was taken as in other ABO$_3$ crystals
like KNbO$_3$ and LiNbO$_3$ \cite{Hewitt61,Peterson68}. For the $^{45}$Sc
ion the antishielding parameter is rather uncertain. However, clearly its
value has to be smaller of about two times in comparison with $^{93}$Nb due
to smaller number of electronic orbitals. Value $\gamma _\infty $ $=-9$
leads to the best agreement with the experimental spectrum.

As an example, the results of $^{45}$Sc complete spectrum calculation in PSN
for several crystal orientations are presented in Fig. \ref{fig7} and
corresponding values of ion shifts as well as their dispersion in Table \ref
{table2}. It is seen excelent agreement between measured and calculated
spectra including frequency region around $\nu _{L}$, where analytical
method gave much worth fit.

The results of calculation in PMN for two crystal-orientations (B II [001]
and [011]) are presented in Fig. \ref{fig8}. The individual contributions
which were related to the Nb shifts in different directions with respect to
oxygen positions are presented as well. The values of ion shifts and their
dispersion obtained by fitting of the experimental NMR spectra are listed in
the Table \ref{table3}. Excellent similarity of ion shifts in PMN and
disordered PSN is obtained. On the other hand, a large relative dispersion
of niobium (or rather oxygen) shift is noted along $<$111$>$ and $<$011$>$
directions. Such a large fluctuations in ion shifts can show their dynamic
character. Actually, we can not expect full motional narrowing of NMR
spectrum, since the frequency shift of the -1/2 $\longleftrightarrow $ 1/2
transition depends on the square of EFG. However, this assumption can be
supported by measuring and analysis of NMR spectra changes with decreasing
temperature. We plan to carry out such analysis in the nearest future. On
the other hand, large dispersion of the quantities is characteristic feature
of relaxors.

Three main comments should be made.

(i) The values presented in Table \ref{table2}-\ref{table3} are the relative
shifts between Nb(Sc) and oxygen ions, i. e., $d_{Nb(Sc)-O}$. Larger
contribution into $d_{Nb(Sc)-O}$ values due to oxygen ions shift rather than
that of Nb(Sc) ions is expected in accordance with x-ray data \cite
{Malibert97}.

(ii) The real values of ion shifts can be somewhat different from those
given in the Tables \ref{table2}-\ref{table3} due to simplifications allowed
in the calculations, including point charges model. However the conclusion
about existence of the displacements, their directions and dispersion is not
related to the model of EFG calculation because it follows directly from
observed shape and width of $^{45}$Sc and $^{93}$Nb NMR lines.

(iii) In the ordered regions the Sc$^{3+}$, Nb$^{5+}$and O$^{2-}$ ions
occupy completely cubic lattice positions, i. e. they are not shifted. On
the contrary, in the disordered regions these all ions are randomly shifted
away from their cubic lattice sites in one of the directions: $<$001$>$, $<$%
011$>$or $<$111$>$. The larger part of ions ($\sim $50$\%$) experiences
shift along $<$001$>$axes.

\section{Discussion}

Evidently, basic question in the understanding of the origin of unusual
physical properties of relaxor ferroelectrics is their local structure,
because at any temperature and even on a micrometric scale the average
symmetry is cubic rather than any lower one (see, for instance, \cite
{Bonneau91,Mathan91} and references therein). Note, that the existence of
diffusive scattering in the powder neutron diffraction patterns observed in
PMN below 340 K was interpreted in Ref. \cite{Mathan92} as the occurrence of
local polar order inside the average paraelectric matrix. However, it is
still a matter of debate whether this polar order is organized as small
ferroelectric domains or as a dipole glass. In contrast to PMN, PSN\ at T $<$
T$_C$ (T$_C$ $\approx $ 350-360 K) undergoes into the $R3m$ rhombohedral
ferroelectric phase \cite{Malibert97,Knight95}. Moreover, this crystal can
be easily obtained in 1:1 Sc/Nb ordered state. It is important to emphasise
that the polar long range order appears not only in perfect 1:1 Sc/Nb
ordered PSN, but even in compositionally disordered crystal \cite{Malibert97}%
, i. e. the compositional disorder does not prevent establishing at least
partly of a long-range order . To our mind this may correspond to mixed
ferro-glass phase. It is important to compare PSN and PMN local structure on
the base of NMR data because many of novel models, which explain the
behaviour of PMN at low temperatures, are based on the existence of 1:1
Mg/Nb ordered micro-regions similar to 1:1 Sc/Nb regions in PSN.

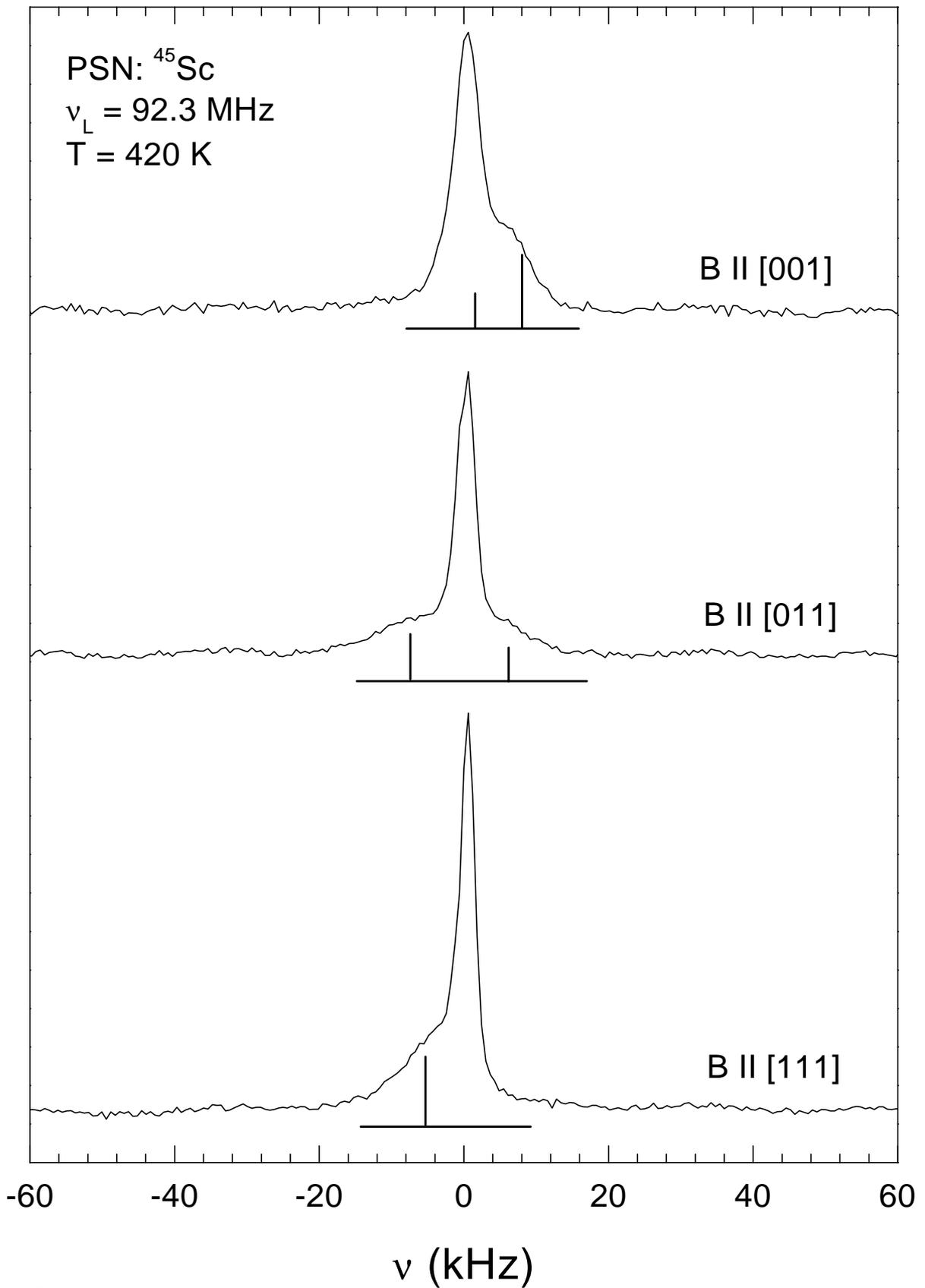
\begin{figure}[tbp]
\caption{$^{45}$Sc NMR spectra in PSN. The diagrams schematecally show 
expected NMR resonances for second order quadrupole contribution with 
EFG axis pointed at $<001>$ directions.}
\label{fig1}
\end{figure}

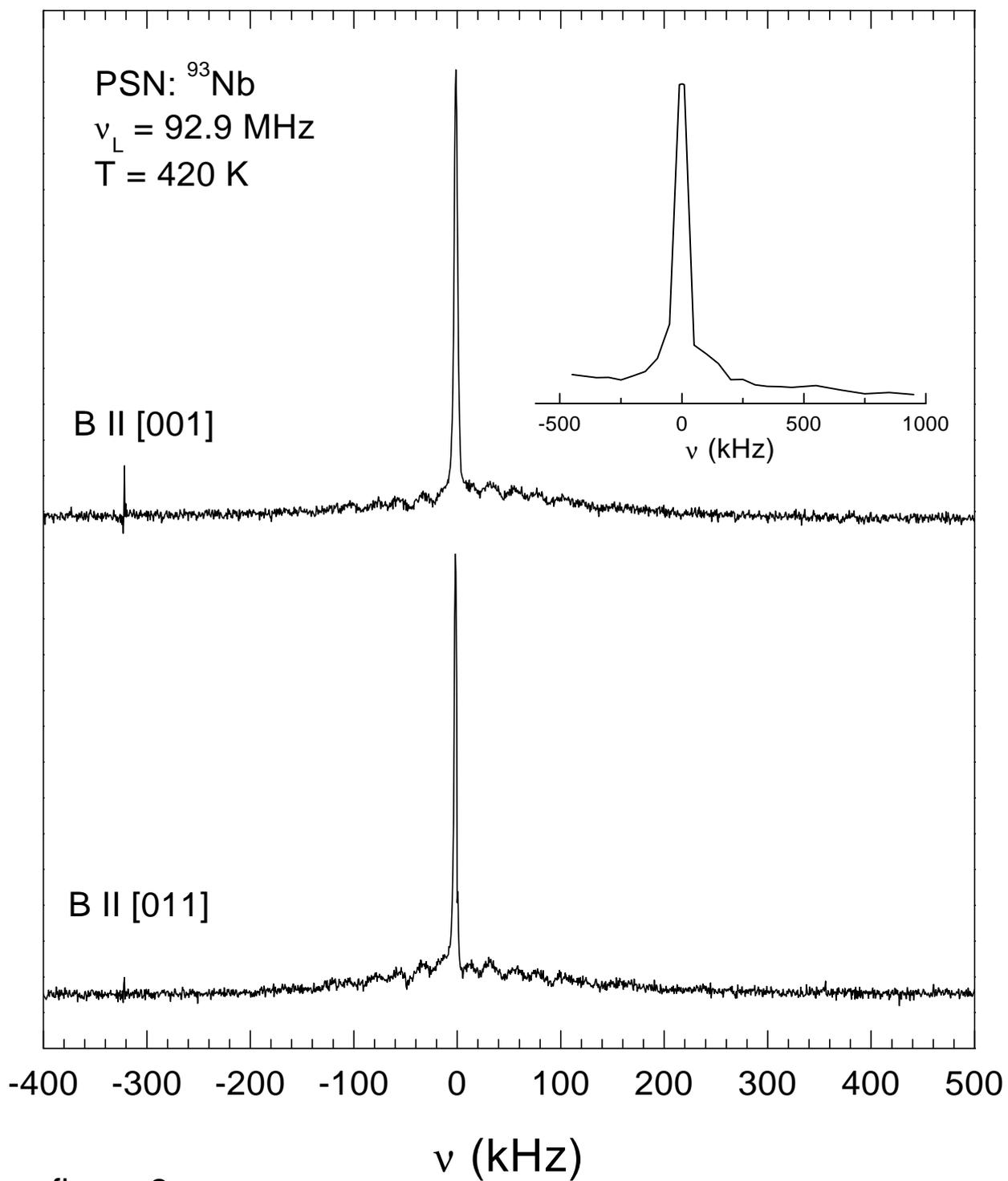
\begin{figure}[tbp]
\caption{$^{93}$Nb NMR spectra in PSN. The inset shows the $^{93}$Nb spectrum 
measured by sweep of the irradiation frequency.}
\label{fig2}
\end{figure}

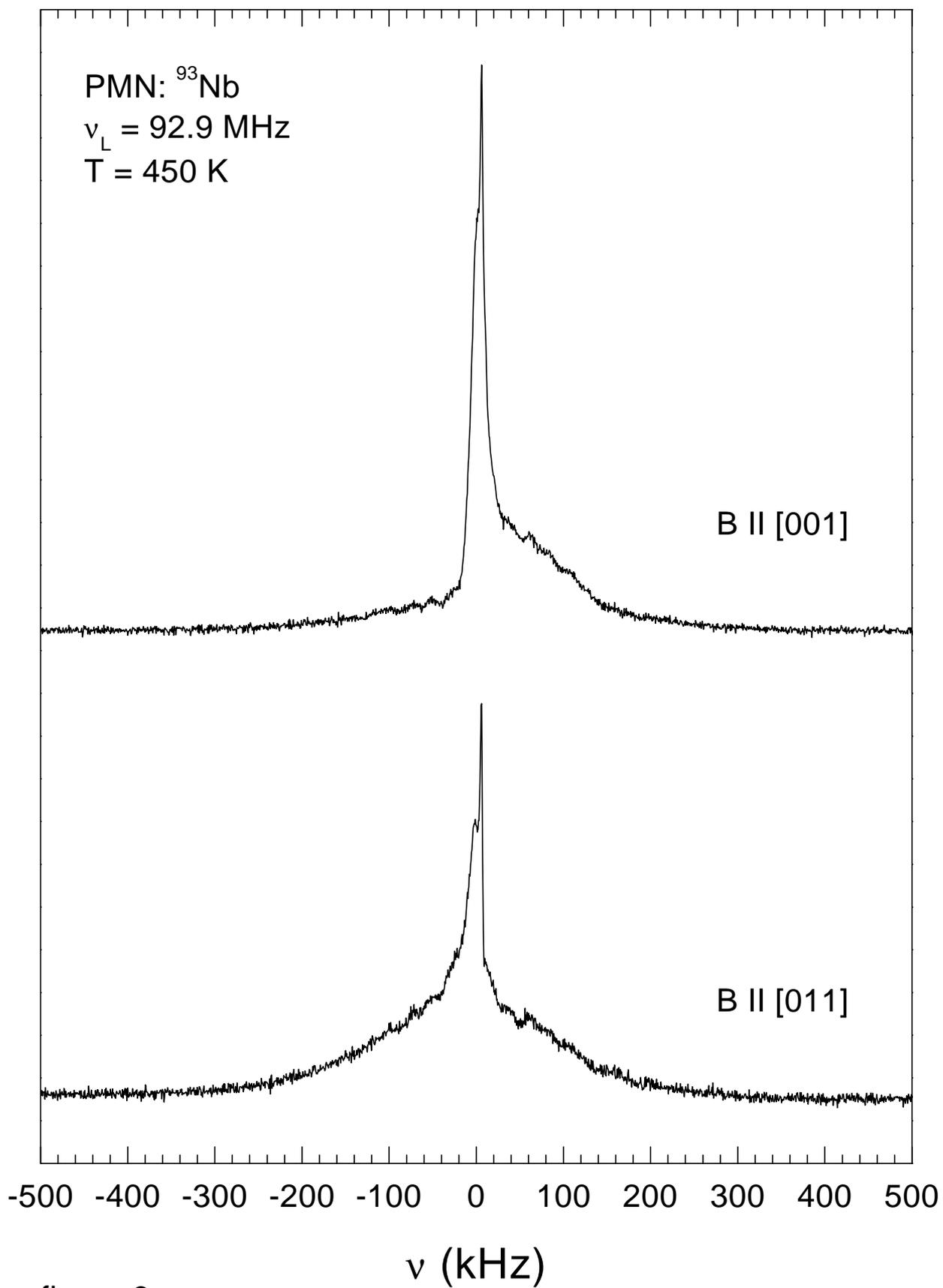
\begin{figure}[tbp]
\caption{$^{93}$Nb NMR spectra in PMN.}
\label{fig3}
\end{figure}

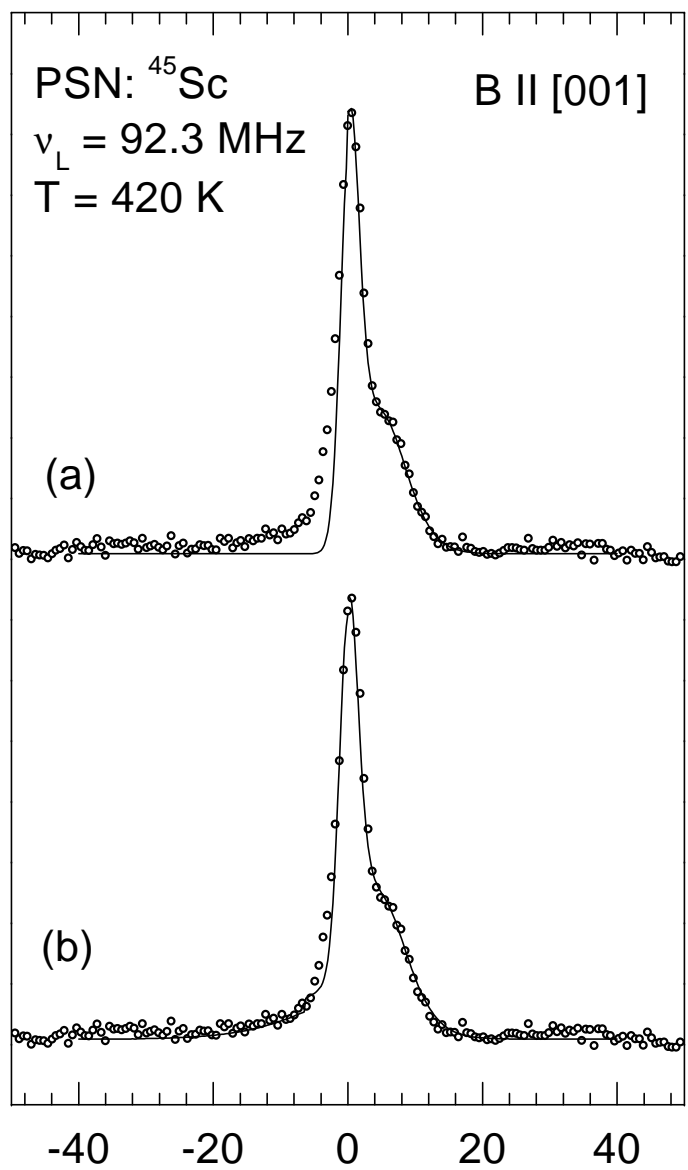
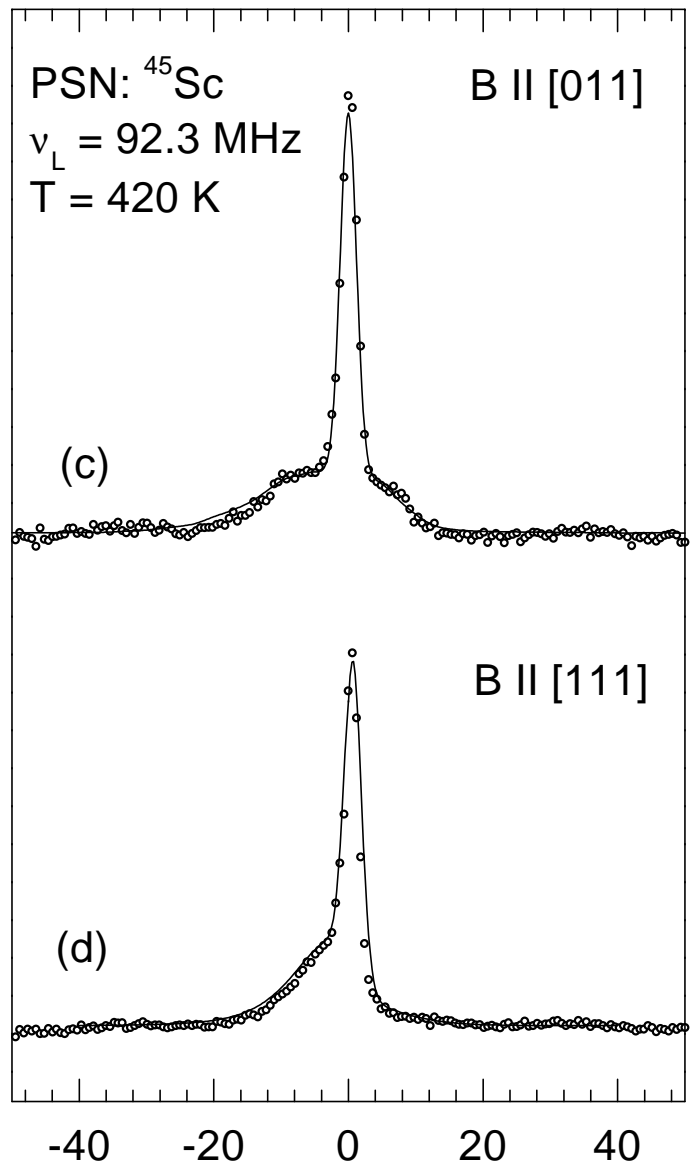
\begin{figure}[tbp]
\caption{Comparison between experimental (points) and calculated (solid
line) $^{45}$Sc NMR line shapes in PSN for a) and b) ${\bf B} \Vert [001]$; 
c) ${\bf B} \Vert [011]$; d) ${\bf B} \Vert [111]$. Calculated
line shapes include contribution from ideal structure regions plus axial EFG
pointed at a) $\left\langle 001\right\rangle $ directions; b), c) and d) $\left\langle
001\right\rangle $, $\left\langle 011\right\rangle $, $\left\langle
111\right\rangle $ directions.}
\label{fig4}
\end{figure}

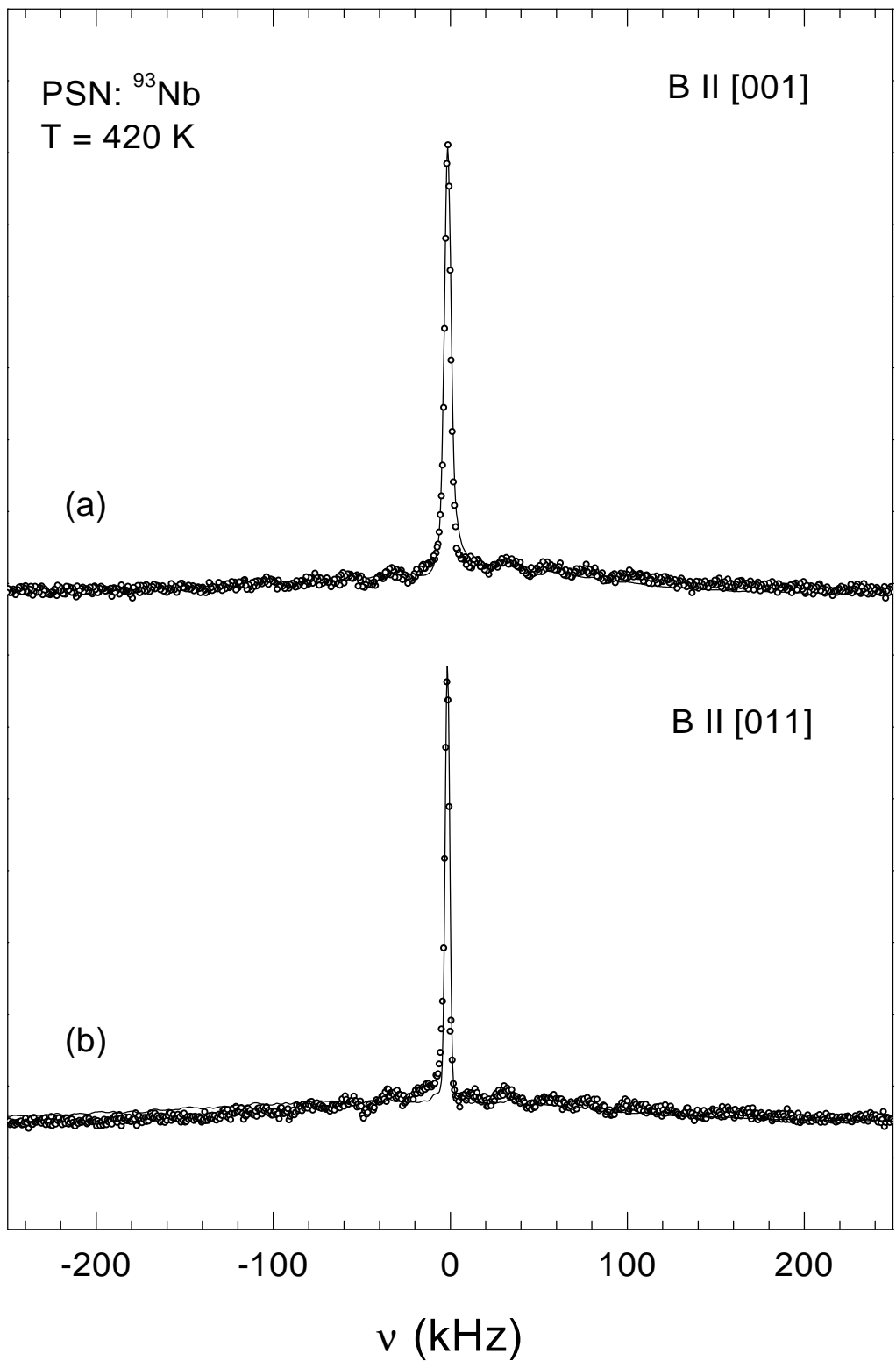
\begin{figure}[tbp]
\caption{Comparison between experimental (points) and calculated (solid
line) $^{93}$Nb NMR line shapes in PSN for a) ${\bf B}\Vert [001]$; b) ${\bf %
B}\Vert [011]$.}
\label{fig5}
\end{figure}

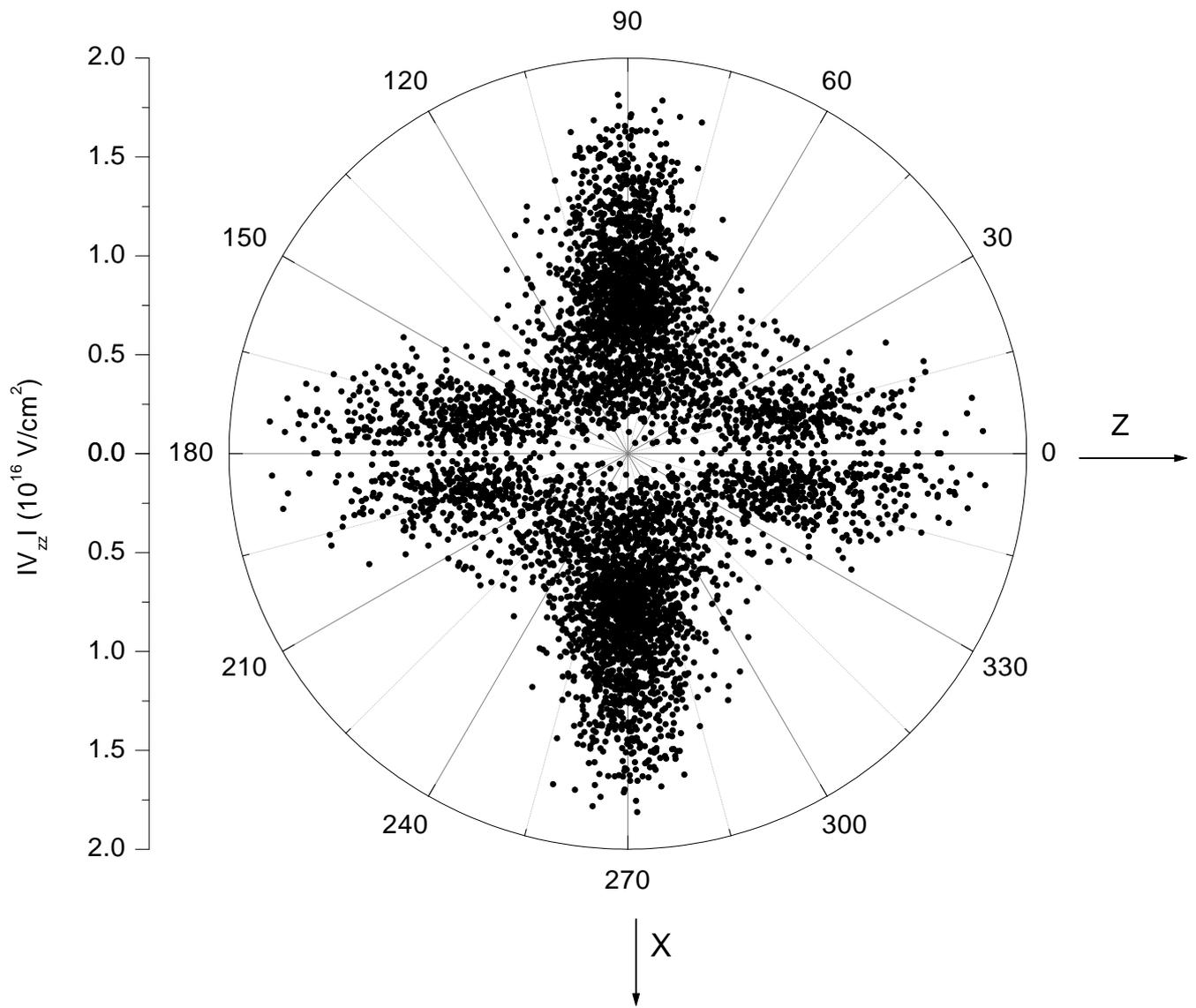
\begin{figure}[tbp]
\caption{Polar plot of the $\mid V_{zz}\mid$ EFG component at Nb sites in PMN
calculated for Mg$^{2+}$ and Nb$^{5+}$ random distribution at lattice points
of ideal ABO$_3$ structure.}
\label{fig6}
\end{figure}

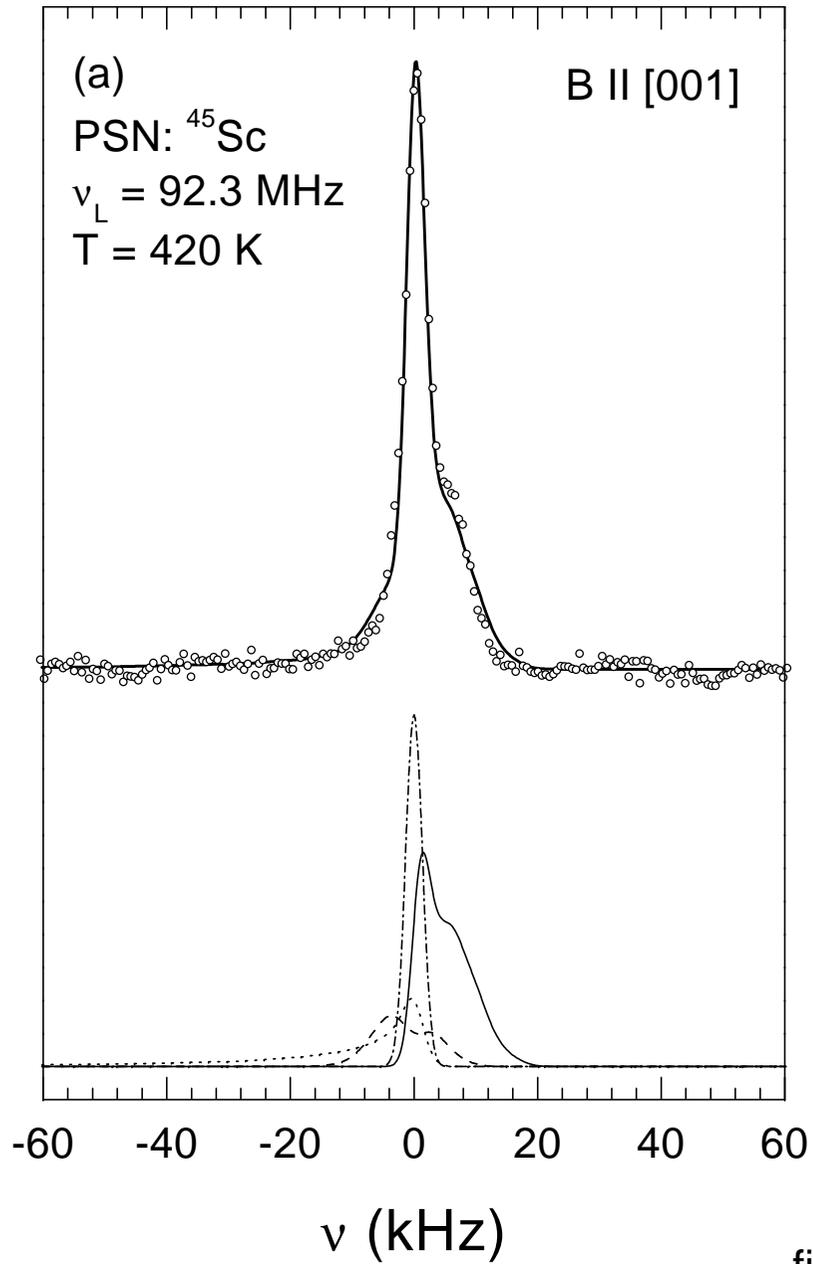
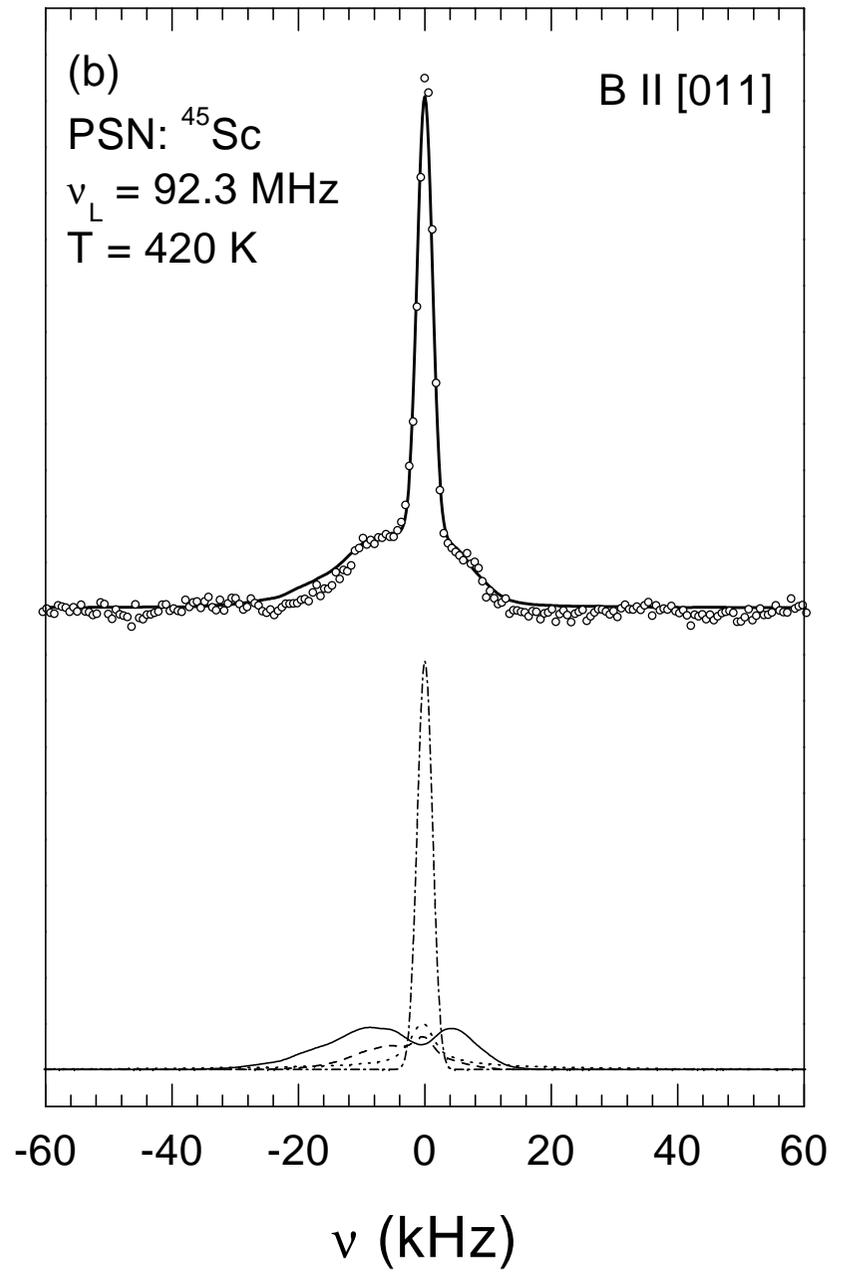
\begin{figure}[tbp]
\caption{Measured and calculated $^{45}$Sc NMR spectra in PSN 
at orientation (a) B II [001] and (b) B II [011]. Separated contributions into NMR spectrum
connected with different type of ion shifts are shown in the bottom panels by: solid line ($%
d_{Sc-O}$ II $<$001$>$), dash line ($d_{Sc-O}$ II $<$011$>$) and dot line ($%
d_{Sc-O}$ II $<$111$>$); contribution from 1:1 Sc/Nb ordered regions is
shown by dash dot line. }
\label{fig7}
\end{figure}

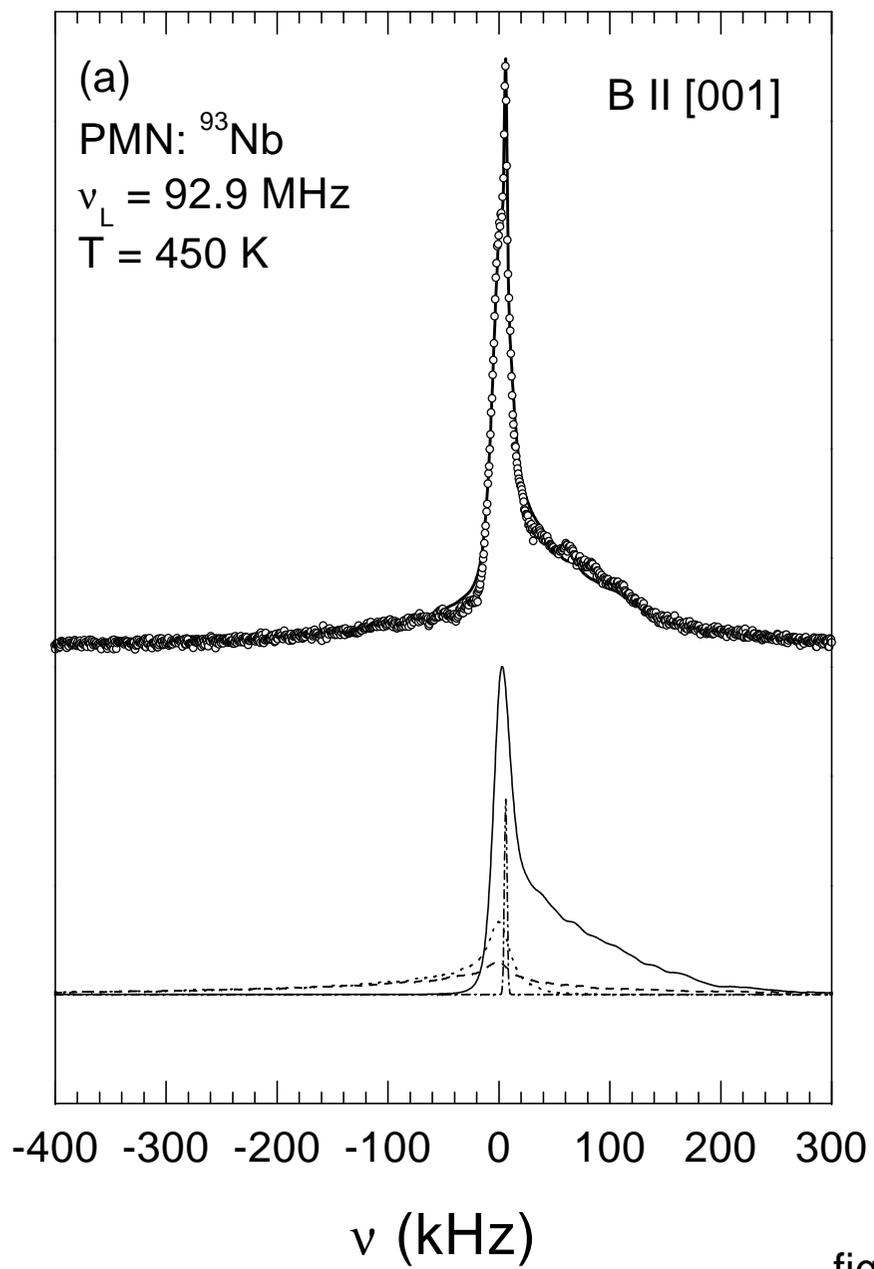
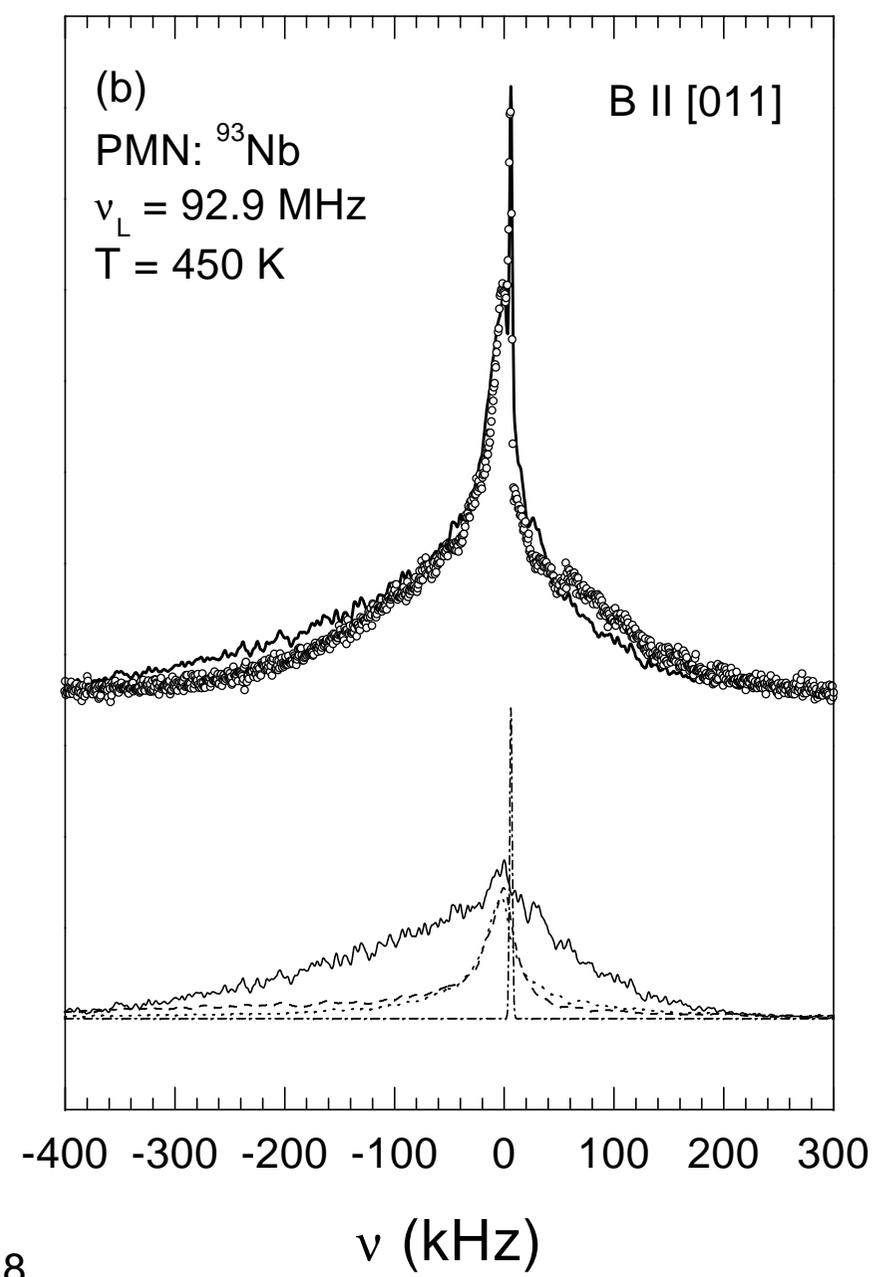
\begin{figure}[tbp]
\caption{Measured and calculated $^{93}$Nb NMR spectra in PMN 
at orientation (a) B II [001] and (b) B II [011]. Separated contributions into NMR spectrum
connected with different type of ion shifts are shown in the bottom panels by: solid line ($%
d_{Nb-O}$ II $<$001$>$), dash line ($d_{Nb-O}$ II $<$011$>$) and dot line ($%
d_{Nb-O}$ II $<$111$>$); contribution from 1:2 Mg/Nb ordered regions is
shown by dash dot line. }
\label{fig8}
\end{figure}

It is clearly seen from Figs. \ref{fig1} and \ref{fig2} that for 1:1 ordered
regions in the PSN distinctive narrow line in both $^{45}$Sc and $^{93}$Nb
NMR spectra are obtained. In these regions Sc and Nb ions are located in
regular cubic lattice sites, where the total electric-field-gradient is zero
or close to zero. The $^{93}$Nb NMR spectrum in PMN also exhibits narrow
component (Fig. \ref{fig3}) but its integral intensity is only of about 1.5
- 2 percent of the total spectrum intensity (Tables \ref{table2},\ref{table3}%
). Evidently, it is reason to ascribe this $^{93}$Nb resonance to the
regular 1:2 Mg/Nb ordered regions of the crystal. Note that this resonance
could not belong to 1:1 Mg/Nb ordered regions because in agreement with high
resolution electron microscopy study \cite{Boulesteix94} the portion of such
ordered regions (if they exists) should be approximately 30$\%$. Thus, if
local ($\sim $ 1.5 - 2.5 nm) 1:1 ordered clusters exist in PMN, the
disagreement between NMR and electron microscopy data can be overcame in
supposition that in these local regions niobium and oxygen ions are shifted
as in completely disordered regions. This fact seems to be strange. Let us
remind that in 1:1 Sc/Nb ordered PSN the Sc and Nb ions are situated in
correct cubic lattice sites. Taking into account this argument, the model
grounded on the alternation of Nb and (Mg$_{2/3}$Nb$_{1/3}$) layers which
also explains the observation of 1/2 $<$111$>$ superlattice diffraction peak 
\cite{Egami98,Chen89} seems to be more convincing. In this model local
strains induced by the charge disbalance between the Nb and Mg/Nb layers can
be locally compensated by oxygen ion displacements which will give the
contribution in the electric field gradient. Additional contribution to EFG
is also due to (Mg$_{2/3}$Nb$_{1/3}$) layer compositional disorder.

Obtained evidence about three possible directions of Nb ions shift
relatively the surrounding ions (so that oxygen and lead ions shift can
contribute also) seems to be very important. Physical nature of this
phenomenon can be related to substitutional disorder in the relaxors A(B$%
_{1/3}^{^{\prime }}$B$_{2/3}^{^{\prime \prime }}$)O$_3$.

We can not exclude also that ion shift in one out of three observed
directions may be the result of ferroelectric phase transition which could
be at Burns temperature T$_d$ = 640 K \cite{Burns}. However with the
temperature decrease strong random fields completely destroy this
ferroelectric long range order because of correlation radius $r_c$ decrease
at T $<$ T$_d$ (here $r_c$ is the correlation radius of the lattice that
should be responsible for the phase transition at T$_d$)\cite{GlFarhi}. As a
result only a small movable polar clusters appear below T$_d$. In such a
picture strong dispersion and V-F law, observed in the temperature
dependence of dielectric permitivity in PMN (see, e.g. \cite{Vill} and ref.
therein) can be considered as manifestation of reentrant phase (like that in
magnetic systems \cite{magsys}) of aforementioned Burns reference phase.

To clear up all aforementioned questions precise measurements of NMR spectra
in the vicinity of the Burns temperature are extremely desirable.

Finally, we would like to point out that proposed methods of NMR line shape
analysis can be applied to other disordered materials where the NMR
frequency shift is a nonlinear function of the random electric field
gradients.

\begin{table}[tbp]
\caption{$^{45}$Sc and $^{93}$Nb quadrupole coupling constants and their
dispertion measured in disordered PSN and PMN at T = 450 K.}
\label{table1}
\begin{tabular}{lcccc}
Compound & $\frac{e^2qQ}h$ (MHz) & $\Delta$ (MHz) & $\mid V_{zz}\mid$ ($10^{20}Vm^{-2}$) & 
Orientation \\ 
&  & & & of V$_{zz}$ \\ \hline
PbMg$_{1/3}$Nb$_{2/3}$O$_3$ &  & & &  \\ 
$^{45}$Sc & 10.6 & 2.9 & 20 & $<$001$>$ \\ 
& 9.1 & 4.9 & 17 & $<$011$>$ \\ 
& 8.8 & 4.9 & 16 & $<$111$>$ \\ 
$^{93}$Nb & 46 & 15 & 68 & $<$001$>$ \\ 
& 42 & 20 & 62 & $<$011$>$ \\ 
& 56 & 27 & 83 & $<$111$>$ \\ 
PbMg$_{1/3}$Nb$_{2/3}$O$_3$ &  & & &  \\ 
$^{93}$Nb & 46 & 24 & 68 & $<$001$>$ \\ 
& 60 & 29 & 89 & $<$011$>$ \\ 
& 36 & 29 & 53 & $<$111$>$%
\end{tabular}
\end{table}

\begin{table}[tbp]
\caption{Parameters of ions shifts in PSN at T = 420 K derived from NMR data.}
\label{table2}
\begin{tabular}{lcccc}
& & \multicolumn{3}{c}{direction of ions shifts} \\
& 1:1 ordered & $<$001$>$ & $<$011$>$ & $<$111$>$ \\ \hline
portion ($\%$) & 30 & 35 & 15 & 18 \\ 
$d_{Sc-O}$ (nm) & 0 & 0.015 & 0.016 & 0.017 \\ 
$\sigma$ (nm) & 0 & 0.001 & 0.001 & 0.001 \\ 
$d_{Nb-O}$ (nm) & 0 & 0.016 & 0.018 & 0.022 \\ 
$\sigma$ (nm) & 0 & 0.005 & 0.005 & 0.005
\end{tabular}
\end{table}

\begin{table}[tbp]
\caption{Parameters of ions shifts in PMN at T = 450 K derived from NMR data.}
\label{table3}
\begin{tabular}{lcccc}
& & \multicolumn{3}{c}{direction of ions shifts} \\
& 1:2 ordered & $<$001$>$ & $<$011$>$ & $<$111$>$ \\ \hline
portion ($\%$) & 2 & 56 & 21 & 21 \\ 
$d_{Nb-O}$ (nm) & 0 & 0.016 & 0.021 & 0.017 \\ 
$\sigma$ (nm) & 0 & 0.005 & 0.01 & 0.01
\end{tabular}
\end{table} 

\begin{references}
\bibitem{Smol54}  G. A. Smolenskii and V. A. Isupov, Dokl. Akad. Nauk SSSR 
{\bf 97}, 653 (1954).

\bibitem{Cross87}  L. E. Cross, Ferroelectrics {\bf 76}, 241 (1987).

\bibitem{Malibert97}  C. Malibert, B. Dkhil, J. M. Kiat, D. Durand, J. F.
Berar, and A. Spasojevic-de Bire, J. Phys.: Condens. Matter, {\bf 9}, 7485
(1997).

\bibitem{Glin}  M. D. Glinchuk, V. V. Laguta, I. P. Bykov, S. Nokhrin, V. P.
Bovtun, M. A. Leschenko, J. Rosa, and L. Jastrabik, J. Appl. Phys. {\bf 81},
3561 (1997).

\bibitem{Blinc}  R. Blinc, J. Dolin\v{s}ek, A. Gregorovi\v{c}, B. Zalar, C.
Filipi\v{c}, Z. Kutnjak, A. Levstik, and R. Pirc, Phys. Rev. Lett. {\bf 83},
424 (1999).

\bibitem{psn}  R. Blinc, A. A. Gregorovi\v{c}, B. Zalar, R. Pirc, V. V.
Laguta, and M. D. Glinchuk, J. Appl. Phys. {\bf 89}, 1349 (2001).

\bibitem{Abraham}  A. Abragam, {\it The Principles of Nuclear Magnetism}
(Clarendon, Oxford, 1961).

\bibitem{GlKon}  M.D. Glinchuk, I.V. Kondakova, Sov. Sol. St. Phys., {\bf 40}%
, 340 (1998).

\bibitem{Das56}  T. P. Das, R. Bersohn, Phys. Rev., {\bf 102}, 733 (1956).

\bibitem{A}  R. E. Cohen, H. Krakauer, Phys. Rev. {\bf B42}, 6416 (1990).

\bibitem{B}  R. I. Eglitis, A. V. Postnikov, and G. Borstel, Phys. Rev. {\bf %
B54}, 2421 (1996).

\bibitem{C}  R. Resta, M. Posternak, and A. Baldereschi, Phys. Rev. Let. 
{\bf 70}, 1010 (1993).

\bibitem{D}  U. V. Waghmare and K. M. Rabe, Phys. Rev. {\bf B55}, 6161
(1997).

\bibitem{Zhong94}  W. Zhong, R. D. King-Smith, and D. Vanderbilt, Phys. Rev.
Lett., {\bf 72}, 3618 (1994).

\bibitem{Hewitt61}  R. R. Hewitt, Phys. Rev., {\bf 121}, 45 (1961).

\bibitem{Viskov66}  A. S. Viskov and Yu. N. Venevcev, Fiz. Tverd. Tela (in
Russian), {\bf 8}, 416 (1966).

\bibitem{Peterson68}  G. E. Peterson and P. M. Bridenbaugh, J. of Chemical
Physics, {\bf 43}, 3402 (1968).

\bibitem{Bonneau91}  P. Bonneau, P. Garnier, G. Calvarin, E. Husson, J. R.
Gavarri, A. W. Hewat, and A. Morell, Journal of Solid State Chemistry, {\bf %
91}, 350 (1991).

\bibitem{Mathan91}  N. -de Mathan, E. Husson, G. Calvarin, J. R. Gavarri, A.
W. Hewat, and A. Morell, J. Phys.: Condens. Matter, {\bf 3}, 8159 (1991).

\bibitem{Mathan92}  de Mathan, E. Husson, and A. Morell, Mater. Res. Bull., 
{\bf 27}, 867 (1992).

\bibitem{Knight95}  K. S. Knight and K. Z. Baba-Kishi, Ferroelectrics, {\bf %
173}, 341 (1995).

\bibitem{Boulesteix94}  C. Boulesteix, F. Varnier, A. Llebaria, and E.
Husson, J. of Solid State Chemistry, {\bf 108}, 141 (1994).

\bibitem{Egami98}  T. Egami, W. Dmowski, S. Teslic, P. K. Davies, I. -W.
Chen, and H. Chen, Ferroelectrics, {\bf 206-207}, 231 (1998).

\bibitem{Chen89}  J. Chen, H. M. Chan, and M. P. Harmer, J. Amer. Cer. Soc.,%
{\bf 72}, 593 (1989).

\bibitem{Burns}  G. Burns and F. H. Dacol, Phys. Rev. B, {\bf 28}, 2527
(1988).

\bibitem{GlFarhi}  M. D. Glinchuk and R. Farhi, J. Phys.: Condens. Matter, 
{\bf 8}, 6985 (1996).

\bibitem{Vill}  M. D. Glinchuk and V. A. Stephanovich, J. Appl. Phys., {\bf %
85}, 1722 (1999).

\bibitem{magsys}  J. Ya. Korenblit and E. F. Shender, Usp. Fiz. Nauk, {\bf %
157}, 267 (1989).
\end{references}
\end{document}